\documentclass[article,showpacs,preprintnumbers]{revtex4}

\usepackage{graphicx}
\usepackage{dcolumn}
\usepackage{bm}

\begin{document}
\newcommand{\bstfile}{aps} 
\draft \title{A new dipole-free sum-over-states expression for the second
hyperpolarizability}

\author{Javier P\'erez-Moreno}
\email{Javier.PerezMoreno@fys.kuleuven.be} \affiliation{Department
of Chemistry, University of Leuven, Celestijnenlaan 200D, B-3001
Leuven, Belgium}

\author{Koen Clays}

\email{Koen.Clays@fys.kuleuven.be} \affiliation{Department of
Chemistry, University of Leuven, Celestijnenlaan 200D, B-3001
Leuven, Belgium,}

\affiliation{Department of Physics and Astronomy, Washington State
University, Pullman, Washington 99164-2814}

\author{Mark G. Kuzyk} \email{kuz@wsu.edu} \affiliation{Department
of Physics and Astronomy, Washington State University, Pullman,
Washington 99164-2814}

\date{\today}

\begin{abstract}

The generalized Thomas-Kuhn sum rules are used to eliminate the
explicit dependence on dipolar terms in the traditional
sum-over-states (SOS) expression for the second hyperpolarizability
to derive a new, yet equivalent, SOS expression. This new
dipole-free expression may be better suited to study the second
hyperpolarizability of non-dipolar systems such as quadrupolar,
octupolar, and dodecapolar structures.  The two expressions lead to
the same fundamental limits of the off-resonance second
hyperpolarizability; and when applied to a particle in a box and a
clipped harmonic oscillator, have the same frequency-dependence.  We
propose that the new dipole-free equation, when used in conjunction
with the standard SOS expression, can be used to develop a
three-state model of the dispersion of the third-order
susceptibility that can be applied to molecules in cases where
normally many more states would have been required.  Furthermore, a
comparison between the two expressions can be used as a convergence
test of molecular orbital calculations when applied to the second
hyperpolarizability. \end{abstract}


\maketitle


\vspace{1em}

\section{Introduction}

The sum-over-states (SOS) expressions have been used for more than
three decades in the study of nonlinear optical phenomena, and are
perhaps the most universally used equations in molecular nonlinear
optics. The sum-over-states expression is obtained from quantum
perturbation theory and is usually expressed in terms of the matrix
elements of the dipole operator, $-ex_{nm}$, and the zero-field
energy eigenvalues, $E_{n}$.\cite{Boyd,OrrWard,KuzBook}\\

The SOS expressions for the first and second hyperpolarizability
derived by Orr and Ward using the method of averages\cite{OrrWard}
are often used because they explicitly eliminate the unphysical
secular terms that are present in other derivations.\cite{Boyd}
These secular-free expressions contain summations over {\em all}
excited states.\\

Finite-state approximations are used to apply the theory to
experimental results. Oudar and Chemla studied the first
hyperpolarizability of nitroanilines by considering only two states,
the ground and the dominant excited state.\cite{OudarChemla}
Although the general validity of this ``two-level'' model has been
questioned, especially in its use for extrapolating measurement
results to zero frequency, the approximation is still widely used in
experimental studies of the nonlinear properties of organic
molecules. \\

Several approaches have been used to develop approximate expressions
for the second-hyperpolarizability in the off-resonance
regime.\cite{kuzdirkfirst,Nakano, Meyers} While such approximations are helpful,
they systematically ignore some of the contributions to the SOS
expression. As our goal is to derive a general expression that is
equivalent to the traditional SOS one, we choose not to make any
assumptions a priori about what type of contributions dominate the
response. Furthermore, including all the possible contribution is
necessary to properly describe the on-resonance behavior, even when
only few states contribute to the response.\cite{TPAmine}\\

In 2005, Kuzyk used the generalized Thomas-Kuhn sum rules to relate
the matrix elements and energies involved in the general Orr and
Ward SOS expression for the first hyperpolarizability, and
introduced a new and compact SOS expression that does not depend
explicitly on dipolar terms.\cite{KuzykNew} Since the Thomas-Kuhn
sum rules are a direct and exact consequence of the Schr\"odinger
equation when the Hamiltonian can be expressed as $H = p^2/2m +
V(r)$, it follows that the new SOS expression is as general as the
original, converges to the same results, and by virtue of its
compactness may be more appropriate for the analysis of certain
nonlinear optical properties.\cite{Fundamental} Indeed, Champagne
and Kirtman used a comparison between the dipole-free and standard
SOS expressions to study the convergence of molecular-orbital
calculations.\cite{champ} In this work, we use the same principle to
derive a compact and general dipole-free expression for the second
hyperpolarizability.

\section{Theory}

While our method can be applied to non-diagonal components of the
second hyperpolarizability, for simplicity we will focus on the
diagonal component.  The SOS expression for the diagonal term of
the second hyperpolarizability $\gamma$ as derived by Orr and Ward
in 1971 is given by:\cite{OrrWard} \begin{equation} \gamma_{xxxx}
(-\omega_{\sigma};\omega_{1},\omega_{2},\omega_{3})=
 e^{4} \left( {\sum_{lmn}^{\infty}}'
\frac{x_{0l}\bar{x}_{lm}\bar{x}_{mn}x_{n0}}{D^{-1}_{lmn}(\omega_{1},\omega_{2},\omega_{3})}-
{\sum_{mn}^{\infty}}'\frac{x_{0m}x_{m0}x_{0n}x_{n0}}{D^{-1}_{mn}(\omega_{1},\omega_{2},\omega_{3})}
\right), \label{dipolefree:eq:gammaxxxx} \end{equation} where $e$ is
the magnitude of the electron charge, $x_{nm}$ the $n,m$ matrix
element of the position operator and $h\omega_{i}$ ($i=1,2,3$) are
the frequencies of the photons with
$\omega_{\sigma}=\omega_{1}+\omega_{2}+\omega_{3}$. The bar operator
is defined as: \begin{equation} \bar{x}_{nm}=\left\{
\begin{array}{cc}
        \Delta x_{n0} \equiv x_{nn}-x_{00} & \mbox{if $n=m$}. \\
        x_{nm} & \mbox{if $n \neq m$}
        \end{array}.
    \right.
\end{equation} The dispersion of $\gamma$ is given by
$D^{-1}_{lmn}(\omega_{1},\omega_{2},\omega_{3})$ and
$D^{-1}_{mn}(\omega_{1},\omega_{2},\omega_{3})$ which are defined as
follows: \begin{eqnarray} \nonumber & &
D_{lmn}(\omega_{1},\omega_{2},\omega_{3}) = \frac{1}{6} \times \\
\nonumber & & \left\{ \frac{1}{(\hbar \Omega_{lg}- \hbar
\omega_{\sigma})(\hbar \Omega_{mg}-\hbar \omega_{1} - \hbar
\omega_{2})(\hbar \Omega_{ng}-\hbar \omega_{1})}\right. \\ \nonumber
& & + \frac{1}{(\hbar \Omega^{*}_{lg}+ \hbar \omega_{3})(\hbar
\Omega_{mg}-\hbar \omega_{1} - \hbar \omega_{2})(\hbar
\Omega_{ng}-\hbar \omega_{1})} \\ \nonumber & & +  \frac{1}{(\hbar
\Omega^{*}_{lg} + \hbar \omega_{1})(\hbar \Omega^{*}_{mg}+\hbar
\omega_{1} + \hbar \omega_{2})(\hbar \Omega_{ng}-\hbar \omega_{3})}
\\ \nonumber & & + \frac{1}{(\hbar \Omega^{*}_{lg} + \hbar
\omega_{1})(\hbar \Omega^{*}_{mg}+\hbar \omega_{1} + \hbar
\omega_{2})(\hbar \Omega^{*}_{ng}+\hbar \omega_{\sigma})} \\
\nonumber & & + \left. \mbox{all six permutations of
$(\omega_{1},\omega_{2},\omega_{3})$ for the above terms} \right\}.
\\ \, \label{eq:dispersion3} \end{eqnarray} \begin{eqnarray}
\nonumber & & D_{mn}(\omega_{1},\omega_{2},\omega_{3}) = \frac{1}{6}
\times \\ \nonumber & & \left\{ \frac{1}{(\hbar \Omega_{mg}- \hbar
\omega_{\sigma})(\hbar \Omega_{mg}-\hbar \omega_{3})(\hbar
\Omega_{ng}-\hbar \omega_{1})}\right. \\ \nonumber & & +
\frac{1}{(\hbar \Omega_{mg}- \hbar \omega_{3})(\hbar
\Omega^{*}_{ng}+ \hbar \omega_{2})(\hbar \Omega_{ng}-\hbar
\omega_{1})} \\ \nonumber & & + \frac{1}{(\hbar \Omega^{*}_{mg} +
\hbar \omega_{\sigma})(\hbar \Omega^{*}_{mg}+\hbar \omega_{3})(\hbar
\Omega^{*}_{ng}+\hbar \omega_{1})} \\ \nonumber & & +
\frac{1}{(\hbar \Omega^{*}_{mg}+ \hbar \omega_{3})(\hbar \Omega_{ng}
- \hbar \omega_{2})(\hbar \Omega^{*}_{ng}+ \hbar \omega_{1})} \\
\nonumber & & + \left. \mbox{all six permutations of
$(\omega_{1},\omega_{2},\omega_{3})$ for the above terms} \right\},
\\ \, \label{eq:dispersion4} \end{eqnarray} where spontaneous decay
is introduced by defining complex energies: \begin{equation} \hbar
\Omega_{n} = E_{n0} - i \Gamma_{n}, \end{equation} where $E_{n0}$ is
the energy different between the $n^{th}$ excited state and the ground
state, and $\frac{\Gamma_{n}}{\hbar}$ is the inverse radiative
lifetime of the $n^{th}$ state.

\subsection{Dipole-free expression for the second
hyperpolarizability}

To obtain a dipole-free expression for the second
hyperpolarizability we begin by separating explicitly dipolar terms
from dipole-free terms in the first term of Eq.
\ref{dipolefree:eq:gammaxxxx}, \begin{eqnarray} \nonumber
{\sum_{n}^{\infty}}' \left( {\sum_{m}^{\infty}}'
\left({\sum_{l}^{\infty}}'
\frac{x_{0l}\bar{x}_{lm}\bar{x}_{mn}x_{n0}}{D_{lmn}^{-1}} \right)
\right) &=& \\ {\sum_{n}^{\infty}}' \frac{(\Delta x_{n0}
x_{0n})^{2}}{D^{-1}_{nnn}} + {\sum_{n}^{\infty}}'{\sum_{m \neq
n}^{\infty}}' \frac{\Delta x_{m0}x_{0m}x_{mn}x_{n0}}{D_{mmn}^{-1}}
&+& \nonumber {\sum_{n}^{\infty}}'{\sum_{l \neq n}^{\infty}}'
\frac{\Delta
x_{n0}x_{0l}x_{ln}x_{n0}}{D_{lnn}^{-1}}+{\sum_{n}^{\infty}}'{\sum_{m
\neq n}^{\infty}}' {\sum_{l \neq m}^{\infty}}' \frac{
x_{0l}x_{lm}x_{mn}x_{n0}}{D_{lmn}^{-1}}.  \\ & & \label{eq:gsplit}
\end{eqnarray} The second term in Eq. \ref{dipolefree:eq:gammaxxxx}
is already dipole-free. \\

It should be noted that for non-dipolar systems (such as octupolar
chromophores), with $\Delta x_{m0} = 0$, only the last term in Eq.
\ref{eq:gsplit} contributes to the second hyperpolarizability. The
generalized Thomas-Kuhn sum rules can be used to obtain a
relationship between the explicitly dipolar terms in terms of only
non-dipolar terms:\cite{KuzykNew} \begin{equation} |x_{k0}|^{2}
\Delta x_{k0} = - {\sum_{n \neq k}}'
\frac{(E_{nk}+E_{n0})}{E_{k0}}x_{0k}x_{kn}x_{n0}.
\label{eq:diagonal} \end{equation} We stress that the only
assumption made in the derivation of Eq. \ref{eq:diagonal} is that
the sum rules hold, which is the case when the unperturbed
Hamiltonian describing the system is conservative.  \\

Substituting Eq. \ref{eq:diagonal} into Eqs. \ref{eq:gsplit} and
\ref{dipolefree:eq:gammaxxxx} yields the dipole-free expression for
the second hyperpolarizability: \begin{eqnarray} \nonumber
\gamma_{xxxx} (-\omega_{\sigma};\omega_{1},\omega_{2},\omega_{3})=
e^{4}{\sum_{n}^{\infty}}'{\sum_{m \neq n}^{\infty}}'{\sum_{l \neq
n}^{\infty}}' \frac{(2E_{m0}-E_{n0})(2E_{l0}-E_{n0})}{E_{n0}^{2}}
\cdot \frac{x_{0m}x_{mn}x_{nl}x_{l0}}{D_{nnn}^{-1}} & & \\ \nonumber
-e^{4}{\sum_{n}^{\infty}}'{\sum_{m \neq n}^{\infty}}'{\sum_{l \neq
m}^{\infty}}' \frac{(2E_{l0}-E_{m0})}{E_{m0}} \cdot
\frac{x_{0l}x_{lm}x_{mn}x_{n0}}{D_{mmn}^{-1}} -
e^{4}{\sum_{n}^{\infty}}'{\sum_{l \neq n}^{\infty}}'{\sum_{m \neq
n}^{\infty}}' \frac{(2E_{m0}-E_{n0})}{E_{n0}} \cdot
\frac{x_{0l}x_{ln}x_{nm}x_{m0}}{D_{lnn}^{-1}} & & \\ \nonumber +
e^{4}{\sum_{n}^{\infty}}'{\sum_{m \neq n}^{\infty}}'{\sum_{l \neq
m}^{\infty}}' \frac{x_{0l}x_{lm}x_{mn}x_{n0}}{D_{lmn}^{-1}} -
e^4{\sum_{mn}^{\infty}}'\frac{x_{0m}x_{m0}x_{0n}x_{n0}}{D^{-1}_{mn}}
& & \\ & & \label{dipolefree:eq:final2}. \end{eqnarray} So, Equation
\ref{dipolefree:eq:final2} is as general as the traditional
sum-over-states expression.\cite{OrrWard}

\section{Applications}

It is useful to compare the convergence between the dipole-free
expression for the second hyperpolarizability (Eq.
\ref{dipolefree:eq:final2}) with the traditional Orr and Ward SOS
expression (Eq. \ref{dipolefree:eq:gammaxxxx}) for various systems.
In this section we will compare these expressions as a function of
wavelength  for two model systems.   Mathematically, both
expressions are equivalent, as long as all excited states of the
system are included in the sum, so this exercise will determine how
many states are required for convergence. Since in practice, the
sum-over-states expressions must be truncated, it is critical to
understand the effect of discarding terms on the nonlinear
susceptibility.  We also apply this new expression to calculate the
fundamental limits of $\gamma$, and show that the results agree with
those obtained using the standard SOS expression.\\

\subsection{Three-level model dipole-free expression: calculation of
the fundamental limit in the off-resonance regime}

We begin by first calculating the fundamental limit of $\gamma$
starting from the dipole-free expression.  The analogous calculation
has already been performed using the traditional Orr and Ward SOS
expression,\cite{KuzTHyper} so we can check whether or not the two
results are the same.  A different set of results would suggest that
the method used in calculating the fundamental limits does not
hold.\\

According to the three-level ansatz,\cite{Fundamental, BCBK,
KuyzkReplies} when near the fundamental limit, only three-levels
contribute to the nonlinear response, Eq. \ref{dipolefree:eq:final2}
becomes: \begin{eqnarray} \nonumber
\frac{\gamma_{xxxx}(-\omega_{\sigma};\omega_{1},\omega_{2},\omega_{3})}{e^{4}}
= \left\{ \frac{D_{111} \, (2E_{20}-E_{10})^{2}}{E_{10}^{2}}+
D_{212} - \frac{(2E_{20}-E_{10})}{E_{10}} \, (D_{211}+D_{112})
\right\}|x_{02}|^{2}|x_{12}|^{2} \\ \nonumber + \left\{
\frac{D_{222} \, (2E_{10}-E_{20})^{2}}{E_{20}^{2}}+ D_{121}-
\frac{(2E_{10}-E_{20})}{E_{20}} \, (D_{122}+D_{221})
\right\}|x_{01}|^{2}|x_{12}|^{2} \\ \nonumber -\left\{
D_{11}|x_{01}|^{4}+(D_{21}+D_{12}) \,
|x_{02}|^{2}|x_{01}|^{2}+D_{22}|x_{02}|^{4} \right\}. \\ \left.
\right. \label{dipolefree:eq:3l} \end{eqnarray} Off-resonance, the
dispersion terms (Eqs. \ref{eq:dispersion3} and
\ref{eq:dispersion4}) simplify to: \begin{equation}
\label{D_lmn_off} D_{lmn}^{off} = 4 \left\{
\frac{1}{E_{l0}E_{m0}E_{n0}} \right\}, \\ \label{D_mn_off}
\end{equation} and \begin{equation} D_{mn}^{off} = 2 \left\{
\frac{1}{E_{m0}^{2}E_{n0}}+\frac{1}{E_{m0}E_{n0}^{2}} \right\},
\end{equation} and the relationships between the first transition
dipole moments can be evaluated from the Thomas-Kuhn sum
rules:\cite{KuzFirst, KuzTHyper, KuzSecond, KuzThird, TPAmine,
Kuzmine, Why} \begin{eqnarray} |x_{02}|^{2} \leq
\frac{E_{10}}{E_{20}} \left[ |x_{01}^{MAX}|^{2}-|x_{01}|^{2} \right
] \label{x_02}\\ |x_{12}|^{2} \leq \frac{E_{10}}{(E_{20}-E_{10})}
\left[ |x_{01}^{MAX}|^{2}+|x_{01}|^{2} \right ],
\label{x_12}\end{eqnarray} with\begin{equation}
|x_{01}^{MAX}|^2=\frac{\hbar^{2}N}{2mE_{10}}, \end{equation} where
$N$ is the number of electrons in the system.\\

Introducing the dimensionless quantities: \begin{eqnarray}
\label{E_ratio} E=\frac{E_{10}}{E_{20}}, \\ \label{x_dim} X =
\frac{|x_{01}|}{|x_{01}^{MAX}|}, \end{eqnarray} the off-resonance
diagonal component of the second hyperpolarizability can be written
as: \begin{equation} \gamma_{xxxx}^{\mbox{\tiny{off}}}= (e
\hbar)^{4} \left(\frac{N}{m} \right)^{2} \frac{1}{E_{10}^{5}} \,
F_{\gamma}(E, X), \label{eq:gamma3exp} \end{equation} where
$F_{\gamma}(E, X)$ is defined by: \begin{equation} F_{\gamma}(E, X)=
- 5(E-1)^{2}(E+1)(E^{2}+E+1)X^{4} - 2(E^{2}-1)E^{3}
X^{2}-(E^{3}+E+3)E^{2}+4. \label{eq:gamma3FX} \end{equation}

\begin{figure}[htp] \centering
\includegraphics[scale=.43]{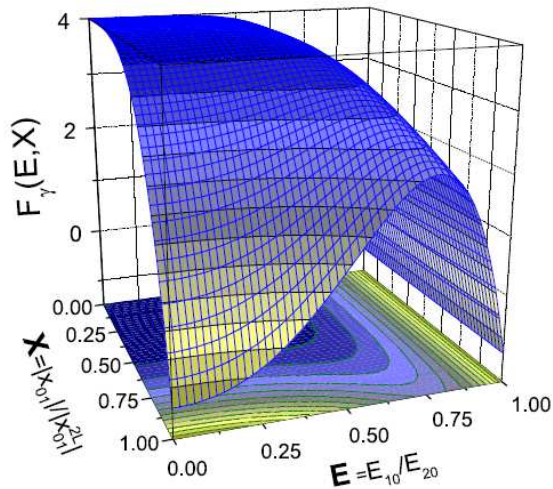}
\caption{\label{fig:gamma3FX} $F_{\gamma}(E, X)$ (as given by Eq.
\ref{eq:gamma3FX}) spanning the full allowable range of the
parameters $X$ and $E$ (defined in Eqs. \ref{E_ratio} and
\ref{x_dim}). The function is optimized when simultaneously $E
\rightarrow 0$ and $X \rightarrow 0$. As $X \rightarrow 1$ and $E
\rightarrow 0$ or $E \rightarrow 1$ the function becomes negative,
corresponding to a negative value of the second
hyperpolarizability.} \end{figure}

The second hyperpolarizability scales as $\gamma \propto N^{2}$, the
square of the number of delocalized electrons, and as $\gamma
\propto \lambda_{max}^{5}$, the fifth power of the wavelength of
maximum absorption. While in a three-level model, the expression for
the first hyperpolarizability, which is analogous to Equation
\ref{eq:gamma3FX}, explicitly separates into a product of a function
of the transition dipole moment $x_{01}$ and excited state energies
(i.e. $F_{\beta}(E,X) = G(X)f(E)$),\cite{Fundamental,Why} this is
not possible for the case of the second hyperpolarizability. To
optimize the second hyperpolarizability the function
$F_{\gamma}(E,X)$ has to be optimized as function of the two
parameters $E$ and $X$. The behavior of the function $F_{\gamma}(E,
X)$ (as given by Eq. \ref{eq:gamma3FX}) as a function of the
parameters $X$ and $E$ is shown in Fig. \ref{fig:gamma3FX}. The
function is maximized when both $E \rightarrow 0$ (i.e. the second
excited state energy level is far away from the first excited energy
level) and $X \rightarrow 0$ (i.e. the oscillator strength is
concentrated in the second transition dipole moment,
$|x_{02}|^{2}$). When the function is optimized we obtain the
quantum limit: \begin{equation}
\gamma_{xxxx}^{\mbox{\tiny{off-max}}} =4\,(e \hbar)^{4}
\left(\frac{N}{m} \right)^{2} \frac{1}{E_{10}^{5}}.
\label{eq:theglimit} \end{equation}  This result agrees with the
quantum limit obtained from the traditional Orr and Ward SOS
expression.\cite{KuzTHyper} \\

Thus, we can conclude that when only three levels contribute to the
nonlinear response, the dipole-free and the traditional SOS
expressions for the second hyperpolarizability - when simplified
using the sum rules - become the same, leading to the same quantum
limits. We should point out that the quantum limits are obtained by
assuming that the response is dominated by the contributions of
three overlapping states, an ansatz that has been extensively verified
numerically using Monte Carlo methods\cite{kuzkuz} as well as
potential energy optimization.\cite{zhoukuzwat} There are no
assumptions about the symmetry
properties of the states. However, this does not imply that symmetry
plays no role in he optimization of the second hyperpolarizability.
The symmetry properties of the system will determine whether or not
the optimal distribution of excited energies and transition dipole
moments can be achieved. Mathematically, symmetries will impose
further constraints on the parameters, which will make $\gamma$ smaller.
We note that the quantum limit is negative for the centrosymmetric system,
and is one-quarter of the positive limit that is obtained for an
asymmetric molecule.\\

\subsection{The particle in a box}

In this section, we test the convergence of the expressions in the
case of two exactly solvable quantum mechanical systems: the
``particle in a box'' and the ``clipped harmonic oscillator''. For
simplicity, we will first perform our calculations in the
off-resonance regime. \\

The unperturbed states that we will use for our calculation of the
second hyperpolarizability are the solutions of the one-dimensional
time-independent Schr\"{o}dinger equation: \begin{equation} -
\frac{\hbar^2}{2m} \frac{\partial^{2} \Psi(x)}{\partial x^{2}} +
V(x) \Psi(x) = E \Psi(x). \end{equation}\\ The potential that
characterizes the particle in a box is zero inside the box of
length $L$ and infinite, otherwise. The solutions are given by:
\begin{equation} \Psi_{k}^{pb}=\sqrt{\frac{2}{L}} \sin \left(
\frac{(k+1)\pi x}{L} \right), \end{equation} with $k=0,1,2,3,\dots$
The corresponding energies are: \begin{equation} E_{k} =
\frac{\hbar^{2}\pi^{2}}{2mL^{2}} (k+1)^{2}, \end{equation} where $m$
is the mass of the particle (in this case the electron mass). These
solutions are substituted into the expressions for the diagonal
component of the second hyperpolarizability (Eqs.
\ref{dipolefree:eq:gammaxxxx} and \ref{dipolefree:eq:final2}) to
study the convergence of both series as a function of number of
excited levels included in the sum.\\

In the off-resonance regime, Eq. \ref{dipolefree:eq:gammaxxxx}
becomes: \begin{equation} \gamma_{xxxx} = 2 e^{4} \left( 2
{\sum_{lmn}^{\infty}}'
\frac{x_{0l}\bar{x}_{lm}\bar{x}_{mn}x_{n0}}{E_{l0}E_{m0}E_{ng}} -
{\sum_{mn}^{\infty}}'x_{0m}^{2}x_{0n}^{2} \left\{
\frac{1}{E_{m0}^{2}E_{n0}} + \frac{1}{E_{n0}^{2}E_{m0}} \right\}
\right), \label{eq:sosoff} \end{equation} and Eq.
\ref{dipolefree:eq:final2} is given by: \begin{eqnarray} \nonumber
\gamma_{xxxx} &=& 2 e^{4} \left( 2 {\sum_{n}^{\infty}}' {\sum_{m
\neq n}^{\infty}}' {\sum_{l \neq n}^{\infty}}' \left\{
\frac{(2E_{m0}-E_{n0})(2E_{l0}-E_{n0})}{E_{n0}^{5}} -
\frac{(2E_{l0}-E_{n0})}{E_{m0}E_{n0}^{3}} \right\}
x_{0m}x_{mn}x_{nl}x_{l0} \right. \\ \nonumber &+& 2
{\sum_{n}^{\infty}}' {\sum_{m \neq n}^{\infty}}' {\sum_{l \neq
m}^{\infty}}' \left\{
\frac{1}{E_{l0}E_{m0}E_{n0}}-\frac{(2E_{l0}-E_{n0})}{E_{m0}^{3}E_{n0}}
\right\} x_{0l}x_{lm}x_{mn}x_{n0} \\ \nonumber &-& \left.
{\sum_{m}^{\infty}}'{\sum_{n}^{\infty}}' \left\{
\frac{1}{E_{m0}^{2}E_{n0}} +\frac{1}{E_{n0}^{2}E_{m0}} \right\}
x_{0m}^{2}x_{0n}^{2} \right). \\ & &\label{eq:dipoleoff}
\end{eqnarray}

The numerical evaluation of the diagonal component of the second
hyperpolarizability as given by Eq. \ref{eq:sosoff} or Eq.
\ref{eq:dipoleoff} respectively, is performed by dividing every
contribution in the sum by the quantum limit (Eq.
\ref{eq:theglimit}). For the particle in a box, a general term in
the sum can be rewritten as: \begin{equation} e^{4}
\frac{x_{0l}x_{lm}x_{mn}x_{n0}}{E_{l0}E_{m0}E_{n0}} =
\frac{3^{5}}{\pi^{4}} \gamma_{xxxx}^{\mbox{\tiny{off-max}}} \cdot
\frac{g^{PB}_{0l}g^{PB}_{lm}g^{PB}_{mn}g^{PB}_{n0}}{f^{PB}_{l0}f^{PB}_{m0}f^{PB}_{n0}},
\end{equation} where we have used defined the following
dimensionless functions: \begin{eqnarray} g^{PB}_{mn} &=&
\int_{0}^{\pi} \sin\left( (m+1) y \right) \cdot y \cdot \sin\left(
(n+1) y \right) dy, \\ f^{PB}_{n0} &=& n(n+2).\end{eqnarray}

\begin{figure}[htp] \centering
\includegraphics[scale=.43]{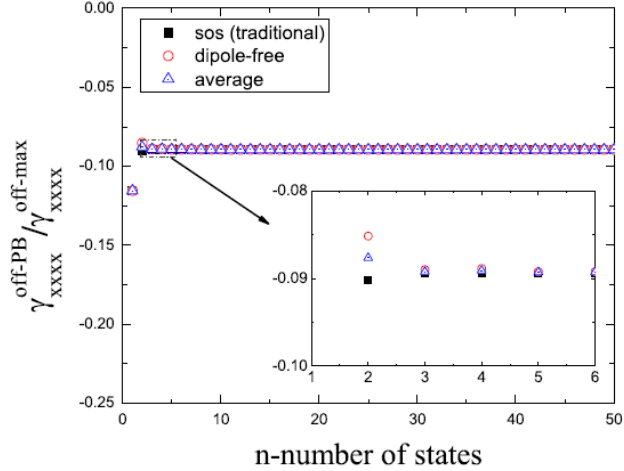}
\caption{\label{fig:particle} Convergence of the normalized SOS
expressions as a function of the number of states in the sum for the
particle in a box. Both the traditional SOS expression (Eq.
\ref{dipolefree:eq:gammaxxxx}) and the dipole-free expression (Eq.
\ref{dipolefree:eq:final2}) converge quickly to the exact value
given by Eq. \ref{finalvalue}. For completeness, we also plot the
normalized average value of the two expressions, which converges
more rapidly as a function of the number of states than the
traditional and dipole-free expressions alone. The inset shows a
mgnified view of the region indicated by the dash boxes.}
\end{figure}

The convergence of the two series in the off-resonance regime is
shown in Fig. \ref{fig:particle}. When including the contribution of
the first 10 states, the relative difference between traditional SOS
expression for the second hyperpolarizability (Eq.
\ref{dipolefree:eq:gammaxxxx}) and the dipole-free expression (Eq.
\ref{dipolefree:eq:final2}) is of the order of $5 \times 10^{-4}$.
With 50 states, the two expressions converge to the same value of:
\begin{equation}
\frac{\gamma^{\mbox{\tiny{off-PB}}}_{xxxx}}{\gamma^{\mbox{\tiny{off-max}}}_{xxxx}}=
-0.08936. \label{finalvalue} \end{equation} Interestingly, the
average value (also shown in Fig. \ref{fig:particle}) is more
accurate when as few as three states are included in the sum. This
is because, for the particular case of the particle in a box, if not
enough states are included in the sums, the traditional SOS
expression tends to underestimate while the dipole-free expressions
tends to overestimate $\gamma$. In this case, with few excited
states, using the average value yields a more reliable estimation of
the second hyperpolarizability.  This same result was found for the
first hyperpolarizability\cite{KuzykNew} and for studies used in
modeling real molecules.\cite{champ}

\subsection{The clipped harmonic oscillator}

Another exactly solvable system is the clipped harmonic oscillator,
whose potential energy function is given by:\cite{Ghatak}
\begin{equation} V(x)=\left\{ \begin{array}{cc}
        \infty & \mbox{if $x<0$}, \\
        \frac{m \omega^{2} x^{2}}{2} & \mbox{if $x \geq 0$}
        \end{array}
    \right.
\end{equation} where $\omega$ has dimensions of frequency. \\

Introducing the dimensionless variable: \begin{equation} \xi
=\sqrt{\frac{m \omega}{\hbar}}x, \end{equation} the solutions are
expressed as: \begin{equation} \Psi_{k}^{cho}=\left\{
    \begin{array}{cc}
    0 & \mbox{if $\xi <0$}, \\
    (2^{2k}(2k+1)!)^{-1/2} \left( \frac{m \omega}{\pi \hbar} \right)^{1/4}
    \exp (-\xi^2/\omega) \, H_{2k+1}(\xi) & \mbox{if $\xi \geq 0$}
    \end{array}
    \right.
\end{equation} where $H_{k}(\xi)$ is the $k^{th}$ order Hermite
Polynomial, and the energies are given by: \begin{equation} E_{k} =
\hbar \omega \left( 2k+\frac{3}{2} \right), \end{equation} with
$k=0,1,2,3,\dots$ \\

For the clipped harmonic oscillator, a general term in the sum can
be rewritten as: \begin{equation} e^{4}
\frac{x_{0l}x_{lm}x_{mn}x_{n0}}{E_{l0}E_{m0}E_{n0}} =
\frac{\gamma_{xxxx}^{\mbox{\tiny{off-CHO}}}}{\pi^{2}} \cdot
\frac{g^{CHO}_{1(2l+1)}g^{CHO}_{(2l+1)(2m+1)}g^{CHO}_{(2m+1)(2n+1)}g^{CHO}_{(2n+1)1}}{l\cdot
m \cdot n}, \end{equation} where we have defined the following
dimensionless function: \begin{equation} g_{mn}^{CHO} = (2^{n-1}
n!)^{-1/2} (2^{m-1} m!)^{-1/2} \int_{0}^{\infty} H_{n}(x) x H_{m}(x)
dx. \end{equation}

\begin{figure}[htp] \centering
\includegraphics[scale=0.43]{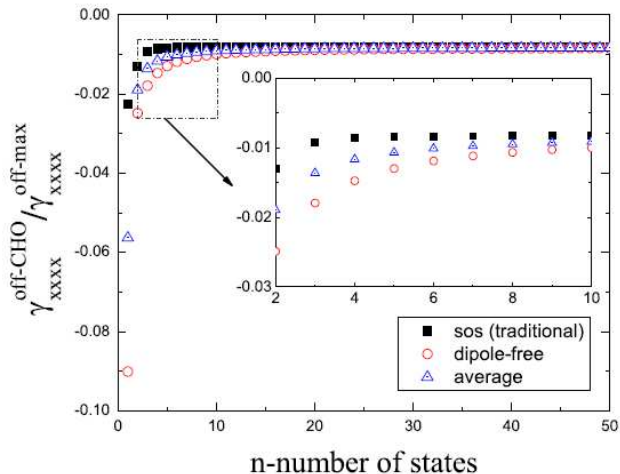}
\caption{\label{fig:cho} Convergence of the normalized SOS
expressions as a function of the number of states in the sum for the
clipped harmonic oscillator. Both the traditional SOS expression
(Eq. \ref{dipolefree:eq:gammaxxxx}) and the dipole-free expression
(Eq. \ref{dipolefree:eq:final2}) converge slowly to the exact value
given by Eq. \ref{finalcho}. For completeness, we also plot the
normalized average value of the two expressions. The inset shows a
magnified view of the region indicated by the dash boxes.}
\end{figure}

The convergence of the two series in the off-resonance regime is
shown in Fig. \ref{fig:cho}. In this case, it takes more terms to
reach convergence than for the particle in a box. For 50 states, the
relative difference between the traditional SOS expression (Eq.
\ref{dipolefree:eq:gammaxxxx}) and the dipole-free expression (Eq.
\ref{dipolefree:eq:final2}) is of the order of $2 \times 10^{-2}$.
The traditional SOS expression converges faster to the final value:
\begin{equation}
\frac{\gamma^{\mbox{\tiny{off-CHO}}}_{xxxx}}{\gamma^{\mbox{\tiny{off-max}}}_{xxxx}}=
-0.00842. \label{finalcho} \end{equation}

\subsection{Dispersion studies using 6 excited levels}

In this section, we investigate the convergence of the dispersion of
the two forms of the second hyperpolarizability.  In particular, we
will treat two separate cases: two photon absorption (TPA) - which
is related to the imaginary part of $\gamma$ and used in a broad
range of applications such as photodynamic cancer therapies and 3D
photolithography,\cite{CancerMaterials, Cancer, PKawata, PCumpston}
and the optical Kerr effect(OKE) - which is related to the real part
of $\gamma$ and widely used in characterizing materials with
potential applications in all-optical switching.\cite{Oke, gates,
shutters} We begin by evaluating Eqs. \ref{eq:dispersion3} and
\ref{eq:dispersion4} and including only the first 6 excited states.
\\

To get the typical qualitative behavior of real large chromophores,
we choose $E_{10}=1eV$. All the linewidths are given by
$\Gamma_{n}=0.1eV$ which is also typical for organic chromophores.
We normalize the second hyperpolarizability by dividing by the
off-resonance limit for $E_{10}=1eV$.\\

\begin{figure}[htp] \centering
\includegraphics[scale=0.43]{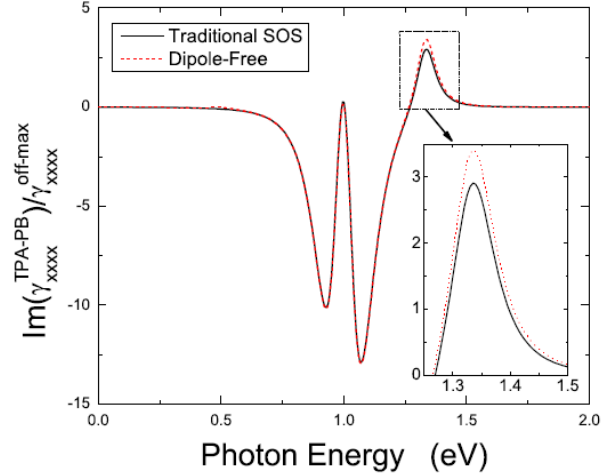}
\caption{\label{fig:disTPAPB} The normalized imaginary part of the
two photon absorption second hyperpolarizability as a function of
the incident photon energy for a 7-level model for the particle in a
box using the traditional sum-over-states and the dipole-free
expressions. The inset shows a magnified view of the region
indicated by the dashed box. $E_{10} = 1eV$.} \end{figure}

Fig. \ref{fig:disTPAPB} shows the dispersion predicted by the
traditional sum-over-states expression and the dipole-free
expression for the imaginary part of the second hyperpolarizability
as s function of the incident photon energy.  Note that $Im[\gamma]$
is related to the two photon absorption cross-section.  The
agreement between the two expressions is excellent everywhere with
the exception of the third resonance (see inset in Fig.
\ref{fig:disTPAPB}), where the two differ by less than 20\%.\\

\begin{figure}[htp] \centering
\includegraphics[scale=0.43]{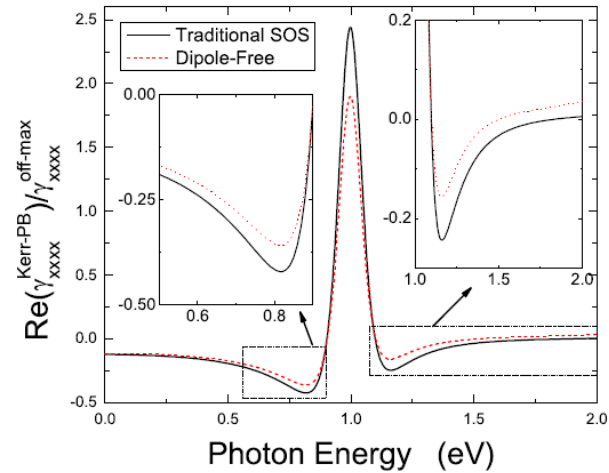}
\caption{\label{fig:disKerrPB} The normalized real part of the Kerr
effect second hyperpolarizability as a function of the incident
photon energy for a 7-level model for the particle in a box using
the traditional sum over states and the dipole-free expressions. The
insets show a magnified view of the region indicated by the dashed
boxes. $E_{10} = 1eV$.} \end{figure}

Next we compare the two expressions for the real part of the second
hyperpolarizability - which is related to the optical Kerr effect -
as a function of the fundamental photon energy for the particle in a
box . The results are plotted in Fig. \ref{fig:disKerrPB}. In this
case, for 6 excited states, the two expressions differ by as much as
a factor of 2 in some regions (see insets). Although the qualitative
behavior is the same, this type of discrepancy should be considered
when experimental data is analyzed using a limited number of terms
in the sum-over-states expressions. \\

\begin{figure}[htp] \centering
\includegraphics[scale=0.43]{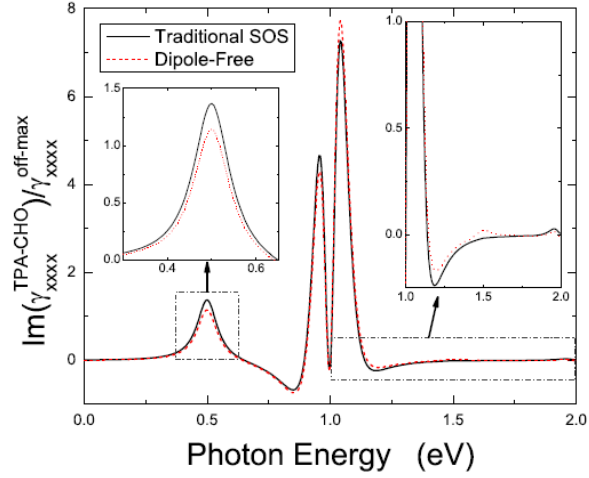}
\caption{\label{fig:disTPACHO} The normalized imaginary part of the
two photon absorption second hyperpolarizability as a function of
the incident photon energy for a 7-level model for the clipped
harmonic oscillator using the traditional sum-over-states and the
dipole-free expressions. The inset shows a magnified view of the
region indicated by the dashed boxes. $E_{10} = 1eV$.} \end{figure}

\begin{figure}[htp] \centering
\includegraphics[scale=0.43]{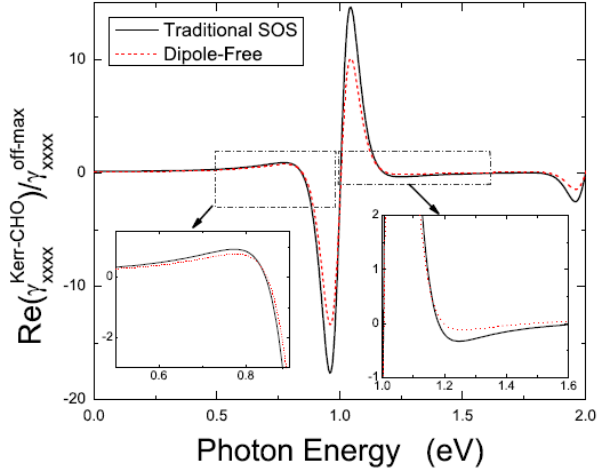}
\caption{\label{fig:dispKerrCHO} The normalized real part of the
Kerr effect second hyperpolarizability as a function of the incident
photon energy for a 7-level model for the clipped harmonic
oscillator using the traditional sum-over-states and the dipole-free
expressions. The insets show a magnified view of the region
indicated by the dashed boxes. $E_{10} = 1eV$.} \end{figure}

Next we consider the dispersion predicted by the traditional
sum-over-states expression and the dipole-free expression for the
imaginary part of the two photon absorption second
hyperpolarizability as a function of the fundamental photon energy for
the clipped harmonic oscillator. The results
are plotted in Fig. \ref{fig:disTPACHO}. The agreement
between the two expressions is good, differing only near the
resonances. Finally, Fig. \ref{fig:dispKerrCHO} shows the dispersion predicted
by the two expressions for the Kerr effect second
hyperpolarizability as a function of the fundamental photon energy
for the clipped harmonic oscillator. For 6 excited states, the two
expressions differ substantially only in the vicinity of a resonance
(see insets).

\section{Convergence of the dipole-free expression}

In this section we will study the convergence of the dipole-free
series expression (Eq. \ref{dipolefree:eq:final2}) as a function of the number of excited
states included in the sum. The convergence is studied as a function
of photon energy for two photon absorption and the Kerr effect,
using the particle in a box and the clipped harmonic oscillator as model
quantum systems. Again, in order to get the typical qualitative
behavior of large real chromophores, we choose $E_{10}=1eV$, and all
the linewidths are given by $\Gamma_{n}=0.1eV$ which is also typical
for organic chromophores. Also, we normalize the second
hyperpolarizability by dividing by the off-resonance limit for
$E_{10}=1eV$.\\

In all the cases it is found that after including about 15
excited states in the sum the expression converges even in the
vicinity of a resonance. In order to get a better understanding of
the convergence behavior of the expressions we will use the particle
in a box model as a test for convergence, once we take into account
symmetry considerations.\\

For clarity, we will look again to the expression for the second
hyperpolarizability that separates explicitly dipolar terms from
dipole-free terms: \begin{eqnarray} \nonumber e^4
{\sum_{n}^{\infty}}' \left( {\sum_{m}^{\infty}}'
\left({\sum_{l}^{\infty}}'
\frac{x_{0l}\bar{x}_{lm}\bar{x}_{mn}x_{n0}}{D_{lmn}^{-1}} \right)
\right) &=& \\ \nonumber e^4{\sum_{n}^{\infty}}' \frac{(\Delta
x_{n0} x_{0n})^{2}}{D^{-1}_{nnn}}+ e^4 {\sum_{n}^{\infty}}'{\sum_{m
\neq n}^{\infty}}' \frac{\Delta
x_{m0}x_{0m}x_{mn}x_{n0}}{D_{mmn}^{-1}} &+&
e^4{\sum_{n}^{\infty}}'{\sum_{l \neq n}^{\infty}}' \frac{\Delta
x_{n0}x_{0l}x_{ln}x_{n0}}{D_{lnn}^{-1}} \\ +
e^4{\sum_{n}^{\infty}}'{\sum_{m \neq n}^{\infty}}' {\sum_{l \neq
m}^{\infty}}' \frac{ x_{0l}x_{lm}x_{mn}x_{n0}}{D_{lmn}^{-1}} &-&
e^4{\sum_{mn}^{\infty}}'\frac{x_{0m}x_{m0}x_{0n}x_{n0}}{D^{-1}_{mn}}.
\label{eq:gammasplit} \end{eqnarray}

For a system whose symmetry demands that all dipole moments vanish,
i.e. $\Delta x_{m0} = 0$ for $m=1,2,3,\cdots$, such as the the
centrosymmetric system of a particle in a box or molecules with
octupolar symmetry (i.e. no dipole moment but non-centrosymmetric),
the first three terms must each vanish.  Numerically, when we use
the dipole-free expression to model such
systems that are centrosymmetric, which demands that all dipole moments vanish, the contributions from the first three terms are precisely zero only when an infinite number of terms are included in the sum.\\

It is useful to consider how many states are needed in order for each of these dipolar terms, written in our new non-dipolar form, to vanish.  We define the partial sums $S_{1}$, $S_{2}$ and $S_{3}$
and $S_{4}$ as:
\begin{eqnarray} S_{1} &=& e^4{\sum_{n}^{\infty}}' \frac{(\Delta
x_{n0} x_{0n})^{2}}{D^{-1}_{nnn}}= e^{4}{\sum_{n}^{\infty}}'{\sum_{m
\neq n}^{\infty}}'{\sum_{l \neq n}^{\infty}}'
\frac{(2E_{m0}-E_{n0})(2E_{l0}-E_{n0})}{E_{n0}^{2}} \cdot
\frac{x_{0m}x_{mn}x_{nl}x_{l0}}{D_{nnn}^{-1}}, \label{eq:S1} \\
S_{2} &=& e^4 {\sum_{n}^{\infty}}'{\sum_{m \neq n}^{\infty}}'
\frac{\Delta x_{m0}x_{0m}x_{mn}x_{n0}}{D_{mmn}^{-1}} =
-e^{4}{\sum_{n}^{\infty}}'{\sum_{m \neq n}^{\infty}}'{\sum_{l \neq
m}^{\infty}}' \frac{(2E_{l0}-E_{m0})}{E_{m0}} \cdot
\frac{x_{0l}x_{lm}x_{mn}x_{n0}}{D_{mmn}^{-1}}, \label{eq:S2} \\
S_{3} &=& e^4{\sum_{n}^{\infty}}'{\sum_{l \neq n}^{\infty}}'
\frac{\Delta x_{n0}x_{0l}x_{ln}x_{n0}}{D_{lnn}^{-1}} = -
e^{4}{\sum_{n}^{\infty}}'{\sum_{l \neq n}^{\infty}}'{\sum_{m \neq
n}^{\infty}}' \frac{(2E_{m0}-E_{n0})}{E_{n0}} \cdot
\frac{x_{0l}x_{ln}x_{nm}x_{m0}}{D_{lnn}^{-1}}, \label{eq:S3} \\
S_{4} &=& e^4{\sum_{n}^{\infty}}'{\sum_{m \neq n}^{\infty}}'
{\sum_{l \neq m}^{\infty}}' \frac{
x_{0l}x_{lm}x_{mn}x_{n0}}{D_{lmn}^{-1}} -
e^4{\sum_{mn}^{\infty}}'\frac{x_{0m}x_{m0}x_{0n}x_{n0}}{D^{-1}_{mn}}.
\label{eq:S4} \end{eqnarray}

Clearly,\, for systems where there is no change in the dipole moment
between the ground and excited states, the dipole-free expression
will converge when simultaneously: $S_{1} \rightarrow 0$,
$S_{2}\rightarrow 0$ and $S_{3} \rightarrow 0$ and $S_{4}
\rightarrow \gamma_{xxxx}
(-\omega_{\sigma};\omega_{1},\omega_{2},\omega_{3})$. We will test
these conditions numerically using the particle in a box as a model
of a centrosymmetric potential with no dipole moment.\\

First we consider two photon absorption and study the convergence of
$S_{1}$ (Eq. \ref{eq:S1}), $S_{2}$ (Eq. \ref{eq:S2}), $S_{3}$ (Eq.
\ref{eq:S3}) and $S_{4}$ (Eq. \ref{eq:S4}) as a function of the
number of excited states for a range of different photon energies.
The results are shown in Fig. \ref{fig:TPApbCombined}. Clearly, for
all cases, few excited states (from 2 to 5) are needed to get good
convergence when away from resonance. While $S_{1}$ also converges
in the resonant regime after 10 excited states are included more
excited states (up to 20) are required for convergence of $S_{2}$
and $S_{3}$ near resonance (see insets). Finally we look at the
convergence of $S_{4}$ (Eq. \ref{eq:S4}), which as we have seen,
converges to the exact value of the second hyperpolarizability for
systems with no permanent dipole moment, such as the particle in a
box. Surprisingly, this expression is shown to converge rapidly as a
function of number of excited states included in the sum, even on
resonance. In fact, it is clear from the plot that only 2 excited
levels might suffice to study the qualitative behavior of the second
hyperpolarizability even close to the resonances.\\

\begin{figure}[htp] \centering

\caption{\label{fig:TPApbCombined} The normalized imaginary part of
the partial sums $S_{1}$ (top-left), $S_{2}$ (top-right), $S_{3}$
(bottom-left) and $S_{4}$ (bottom-right) for two photon absorption
as a function of the incident photon energy. The number of states
included in the sum is shown in the figure legends. The particle in
a box is used as a model centrosymmetric system with vanishing
dipole moments, and therefore as the sums converge we must have
$S_{1} \rightarrow 0$, $S_{2} \rightarrow 0$, $S_{3} \rightarrow 0$
and $Im(S_{4})=Im(\gamma_{xxxx}^{\mbox{\tiny{TPA-PB}}})$. The insets
show a magnified view of the regions indicated by the dashed boxes.}
\end{figure}

\begin{figure}[htp] \centering
\includegraphics[scale=0.65]{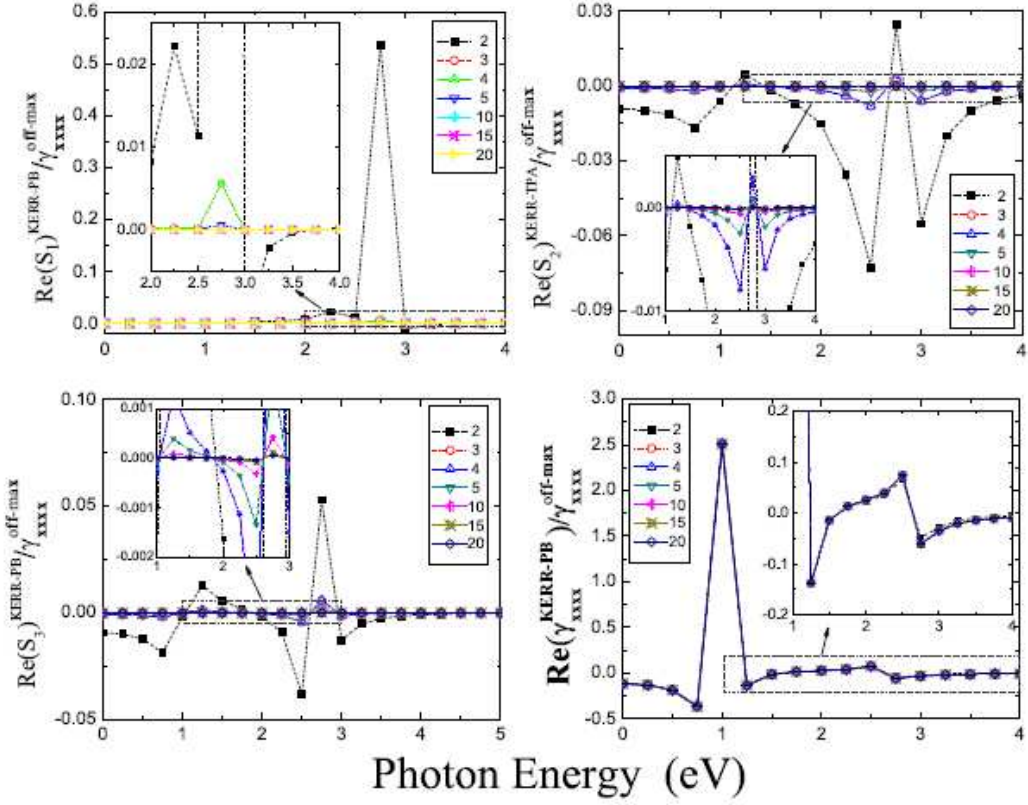}
\caption{\label{fig:KERRpbCombined} The normalized real part of the
partial sums $S_{1}$ (top-left), $S_{2}$ (top-right), $S_{3}$
(bottom-left) and $S_{4}$ (bottom-right) for the Kerr effect second
hyperpolarizability as a function of the incident photon energy. The
number of states included in the sum is shown in the figure legends.
The particle in a box is used as a model centrosymmetric system with
vanishing dipole moments, and therefore as the sums converge we must
have $S_{1} \rightarrow 0$, $S_{2} \rightarrow 0$, $S_{3}
\rightarrow 0$ and
$Re(S_{4})=Re(\gamma_{xxxx}^{\mbox{\tiny{KERR-PB}}})$. The insets
show a magnified view of the region indicated by the dashed boxes.}
\end{figure}

Similarly, we next consider the Kerr effect and study the
convergence of $S_{1}$ (Eq. \ref{eq:S1}), $S_{2}$ (Eq. \ref{eq:S2})
and $S_{3}$ (Eq. \ref{eq:S3}) and $S_{4}$ (Eq. \ref{eq:S4}) as a
function of the number of excited states for a range of different
photon energies. The results are shown in Fig.
\ref{fig:KERRpbCombined}. Again, few excited levels are needed for
convergence away from resonance, although more excited states (up to
20) are required for convergence near resonance. As in the case of
the two photon absorption second hyperpolarizability, $S_{4}$ (Eq.
\ref{eq:S4}), which is the only terms that contributes to the second
hyperpolarizability for the centrosymmetric potential given by the
particle in a box is shown to converge rapidly as a function of
number of excited states included in the sum, even in the resonant
regime.\\

\begin{figure}[htp] \centering
\includegraphics[scale=0.5]{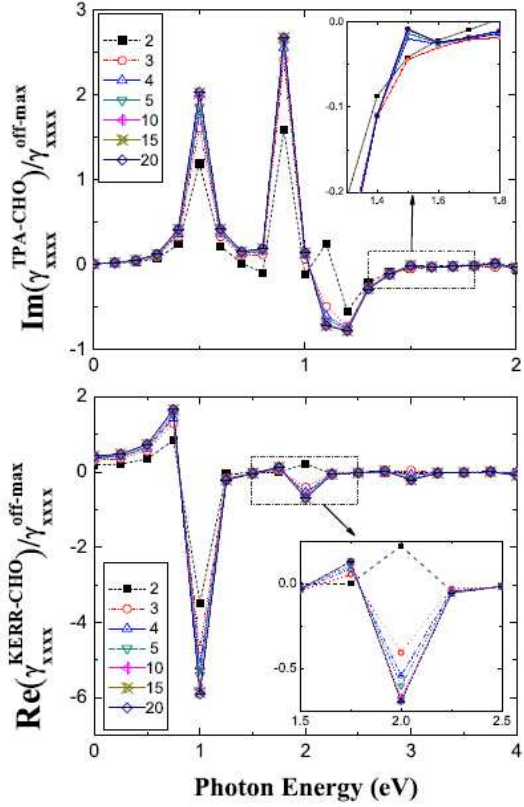}
\caption{\label{fig:choCombined} The normalized imaginary part of
the two photon absorption second hyperpolarizability (top) and the
normalized real part of the Kerr effect second
hyperpolarizability(bottom) as a function of the incident photon
energy. The number of states included in the sum is shown in the
figure captions. The clipped harmonic oscillator is used as a model,
with $E_{10} = 1eV$. The insets show a magnified view of the regions
indicated by the dashed boxes.} \end{figure}

For completeness, we also study the convergence of the expressions
for the two photon absorption process and the Kerr effect using the
clipped harmonic oscillator as a quantum model. As shown by the
results in Fig. \ref{fig:choCombined} the expressions do not
converge as rapidly as did the particle in a box; but when 15
excited states are included in the sum, the spectral features do not
change quantitatively.\\

All of our results show that the dipole-free expression
converges even in the resonant regime when enough excited
states (up to 20) are included in the expression.  We have also shown
that for systems with no dipole moments (such as octupolar
molecules or the particle in a box) the dipole-free expression for
the second hyperpolarizability collapses to Eq. \ref{eq:S4} (which
we will call the {\em reduced} dipole-free expression), since all the
other terms vanish. In this case, the reduced dipole-free expression converges when only a few excited states are included in the sum.

\section{Conclusions}

We have developed an expression that eliminates the explicit dependence on dipolar terms but is physically equivalent to the
traditional SOS expression for the second hyperpolarizability. The equivalence
between the dipole-free and the traditional SOS expressions is
demonstrated by calculating the quantum limits and studying the
convergence of the series with the exact wavefunctions of two
quantum systems: the particle in a box and the clipped harmonic
oscillator. In both cases, when a large number of states is
included, the two expressions are identical.  However, the average
of the two expressions converges faster than the individual
expressions.\\

Since the average between the two expressions appears to be a better
approximation to molecular dispersion, the average may make it
possible to use limited-state models when interpreting experimental
dispersion data.  Since accurate measurements of transition moments
between excited states are difficult and tedious, the averaged
second hyperpolarizability can be a useful tool for modeling the
second hyperpolarizability when only limited information is
available
about the excited states of a particular system.\\

To test the convergence between the two expressions, we have
evaluated them in the resonant regime in two model systems: the
particle in a box - which is a symmetric potential with no change in
dipole moment; and the clipped harmonic oscillator - an asymmetric
potential.  This allows us to determine the role of symmetry.  In
both cases, we study the dispersion of the second
hyperpolarizability near resonance, where - based on the different
energy denominators - one would expect the differences between the
two expressions to be the least consistent. The reduced dipole-free
expression has been introduced for systems with no dipole moment.
Such an expression might be most appropriate when experimental
results are interpreted since it requires the minimum number of
molecular parameters.\\

In conclusion, the dipole-free expression is an alternative to the
traditional SOS expression that increases the theoretical pallet
available to quantum chemists.  It is more direct in certain
theoretical problems such as its application to the derivation of a
more-rigorous calculation of the fundamental limits of the
third-order susceptibility.  It provides a tool to assess the
convergence of truncated SOS calculations, can be used to determine
the accuracy of molecular-orbital calculations of nonlinear
susceptibilities, and can be used to refine limited-state models to
interpret experimental results.  And, it may be more naturally
applicable to the analysis of specific systems such as octupolar
structures.\cite{ZyssJCP, ZyssCR, VerbiestJACS, VerbiestOL, Bidault,
Ledoux-Rak, Ratera, LeBozec, Zyss1993} \\

{\bf Acknowledgements: } JPM acknowledges the Fund for Scientific
Research Flanders (FWO) and the support from the Division of
Molecular and Nanomaterials at the Department of Chemistry in
KULeuven. MGK thanks the National Science Foundation (ECS-0354736)
and Wright Paterson Air Force Base for generously supporting this
work.

\bibliographystyle{\bstfile}

\end{document}